\documentstyle[aps]{revtex}
\begin{document}
\title{Tunability of acoustic spectral gaps and  transmission\\ in periodically 
stubbed 
waveguides }


\author{X. F. Wang$^1$, M. S. Kushwaha$^2$, and P. Vasilopoulos$^1$\\
\ \\}

\address{
$^1$Department of Physics, Concordia University,\\ 1455 de Maisonneuve Blvd. 
West, 
Montreal, Quebec, Canada H3G 1M8\\
\ \\
$^2$Institute of Physics, University of Puebla, P.O. Box J-45, Puebla 72570, Mexico}
\address{}
\address{\mbox{}}
\address{\parbox{14cm}{\rm \mbox{}\mbox{}\mbox{}
A theoretical investigation is made of  acoustic wave propagation in a 
periodically  stubbed waveguide. 
In general the 
waveguide segments and stubs are made of different materials. 
The acoustic wave in such a system has two 
independent polarizations: {\it out-of-plane} and {\it in-plane} modes. 
The band structure and transmission spectrum 
is studied for diverse geometries
using  a simple and efficient version of the 
transfer-matrix method.
For the {\it same material} between the waveguide and {\it symmetric} stubs 
the width of some gaps can change, upon varying the stub length or width,
by more than one order of magnitude. 
A further modulation can be achieved for  {\it different material} between the 
stubs and 
the main waveguide or if the stubs are
{\it asymmetric}.
The gaps in the band structure of an infinitely long system   correspond
to those in the transmission spectrum of the same system but with 
{\it finite} number $n$ of units. 
For $n$ finite i) there exist pseudogaps  that gradually turn into 
complete gaps with increasing $n$, and ii) the introduction of defects 
gives rise to states in the gaps and leads to transmission resonances.
\\}}
\address{\mbox{}}
\address{\parbox{14cm}{
}}
\maketitle

\newpage

\section{INTRODUCTION}

The term ``band-gap engineering" is well known from decades of research in 
semiconductors. The  recently discovered  periodic dielectric structures, which
exhibit a photonic band gap (PBG) analogous to the electronic 
band gap in semiconductors, have attracted considerable attention due
to many interesting phenomena and potential applications emerging from them, 
such as the control of spontaneous emission of radiation, zero-threshold lasing, and the 
sharp bending of light\cite{souk}. 

It did not take long before the study of  PBG materials, involving light waves, 
led to  analogous studies in other systems involving elastic/acoustic waves, e.g., 
the phononic crystals\cite{kush} or other periodic acoustic 
composites\cite{sig}. The phononic crystals\cite{kush} have drawn comparatively greater 
attention, both theoretically\cite{kush1} and experimentally\cite{mon,rob}. In 
analogy with PBG  crystals, the emphasis in phononic crystals has been on the 
occurrence of complete acoustic gaps within which the sound, vibrations, and 
phonons are all forbidden. This is of interest for applications such as 
ultrasonic filters, noise control, and improvement in the design of transducers, 
as well as for fundamental physics concerned with the Anderson localization of 
sound and vibrations\cite{wea}. 

The purpose of this paper is to study  the acoustic band 
structure and transmission spectrum in a periodically modulated 
quasi-one-dimensional waveguide, as depicted in Fig. 1.  The system has  
a finite 
(infinite) extension along the $y$ ($z$)
direction and is periodically modulated,  along 
the $x$ direction, by the addition of double stubs, in general asymmetric,
 with different elastic properties than those of the main waveguide. 
The motivation stems  from  recent studies  with interesting results pertinent 
to electronic\cite{aki}  and photonic\cite{aki1} waveguides modulated 
in the same fashion. Using the transfer-matrix technique  
we demonstrate the tunability of the acoustic band gaps as a function 
of various parameters of the system, e.g, the length and/or width of the stubs.

The rest of the paper is organized as follows. In Sec. II we introduce the 
formalism 
for studying the wave propagation in the allowed polarization and
present the necessary details of the transfer-matrix method\cite{wu}.
In Sec.III we present several illustrative numerical results on the band 
structure and 
transmission spectrum under various material and geometric conditions. The final 
section contains the concluding remarks.

\section{FORMALISM}

This section is divided into two parts. First we embark on the polarization 
pertinent to  wave propagation in a two-dimensional (2D) system. 
Then we present the
transfer-matrix method for quite a general geometry of the
unit cell of acoustic stub tuners shown in Fig. 2.

\subsection{Polarization of the wave} 

We start with the general equation of propagation of harmonic
acoustic waves
in an  isotropic three-dimensional (3D) homogeneous medium

\begin{equation}
(\lambda+\mu)\vec{\nabla}(\vec{\nabla}{\bf{\cdot}}\vec{u})
+\mu \nabla^2 \vec{u} + \rho \,\omega^2 \vec{u} = 0,
\end{equation}
where $\rho$ is the mass density, $\omega$ the angular frequency, and $\lambda$ 
and $\mu$ the Lame coefficients. The longitudinal, $v_l$, and transverse, 
$v_t$, speed of sound are defined in terms of the Lame coefficients: 
$v_l=\sqrt{(\lambda+2\mu)/\rho}$ and $v_t=\sqrt {\mu/\rho}$. For a 2D system at 
hand the displacement vector $\vec{u}$ is independent of the z coordinate and 
one can take $\partial_z\equiv \partial/\partial z=0$. Then Eq. (1) can be written as 

\begin{equation} 
(\lambda +\mu) \vec{\nabla}_p (\partial_x u_x +\partial _y u_y)+
\mu\nabla_p^2(u_x\hat{i}+u_y\hat{j}+u_z\hat{k})+
\rho \omega^2 (u_x\hat{i}+u_y\hat{j}+u_z\hat{k}) = 0,
\end{equation}
where $\vec{\nabla}_p=\hat{i}\partial_x+\hat{j}\partial_y$, and $\hat{i}$, 
$\hat{j}$, and $\hat{k}$ are the unit vectors along the x, y, and z axes 
respectively. This 
equation is equivalent to the 
independent equations

\begin{equation}
(\nabla_p^2+k_t^2)u_z = 0,
\end{equation}
with $k_t=\omega/v_t$, and 

\begin{equation}
(\lambda+\mu)\vec{\nabla}_p (\vec{\nabla}_p {\bf{\cdot}}\vec{u}_p)+
\mu \nabla^2_p\vec{u}_p + \rho\, \omega^2 \vec{u}_p = 0.
\end{equation}
Here the subscript $p$ is assigned to the quantities which qualify only in the 
x-y plane. It is thus quite reasonably understandable that a 2D system can 
support two independent modes: the {\it out-of-plane} modes and the {\it 
in-plane} modes, described, respectively, by Eqs. (3) and (4).   
Equation (4) can be further simplified as follows. We write

\begin{equation}
\vec{u}_p = \vec{\nabla}_p \phi + \vec{\nabla}_p\times\vec{\psi}
\end{equation}
with $\vec{\psi}\equiv (0, 0, \psi)$. Then  Eq. (4) further splits into the equations

\begin{equation}
(\nabla^2_p+k_l^2)\phi=0,
\end{equation} 
with $k_l=\omega/v_l$, and
\begin{equation}
(\nabla^2_p+k_t^2)\psi=0.
\end{equation}
Interestingly,  
Eq. (7), which describes the transverse in-plane vibrations, has formally the 
same structure as the one for the out-of-plane vibrations, Eq. (3). 
Also, Eqs. (3), (6) and (7)
are formally identical to 
the scalar wave equation for the TE polarization in   photonic 
crystals\cite{aki1}.  

It should be pointed out that the splitting of Eq. (4) into Eqs. (6) and (7),
is valid only for  a homogeneous medium. For an inhomogeneous medium, when
$\lambda$ and $\mu$ are functions of position, this is no longer possible
\cite{kush3}. If the system is piecewise homogeneous, as the one we
consider in Sec. III B, problems arise in the application of the boundary 
conditions at the interface of different regions\cite{lan} that make the 
separation of the in-plane modes in pure longitudinal and transverse ones
impossible.

\subsection{Transfer-matrix technique}

For the sake of generality, we start with a crossbar-like geometry of a single 
unit cell, as shown in Fig. 2.   
The origin of the Cartesian coordinates is at the uniaxial 
line intersecting perpendicularly the left arm of the stub of  width $b$ and 
length  $h$. The center of the asymmetric stub lies at ($x=b/2, d$). We denote the 
width of the left (right) waveguide segments by $c$ ($a)$ and take the $x$ 
axis parallel to the direction of propagation. We are interested in the 
solution of the wave equation for the out-of-plane vibrations in the form

\begin{equation}
\nabla^2\phi+k^2\phi = 0,
\end{equation}
where $\phi\equiv u_z$, $k\equiv k_t$, and $\nabla^2\equiv \nabla^2_p$. It is 
very important to note that we consider, for the sake of simplicity, that the 
outer medium containing the said acoustic device is made up of some high-density, 
infinitely rigid material. The resulting situation is equivalent to that 
attained in the case of  similar electronic devices surrounded by infinitely 
repulsive walls\cite{aki}. 

In order to solve the scalar equation 
and  describe the system we use the same transfer-matrix method that was
employed 
in the study of electronic \cite{wu,aki} and photonic \cite{aki1} tuners. The method 
relates the incoming to the outgoing wave across the stub for arbitrary initial 
conditions. Inside the waveguide segments, since the solution must vanish on the walls, 
the $y$ dependence is  $\sin(n\pi (y+c/2)/c)$ for the left segment, for 
example. Here the integer $n$ defines the number of modes in the 
respective waveguide. When the two segments connected with the stub 
have different widths and elastic properties, the respective solutions are given 
by

\begin{equation}
\phi_1=\sum_m \left [ c_{1m}e^{i\beta_mx}+\overline{c}_{1m}e^{-i\beta_mx}
 \right ]
\sin \left (c_m (y+c/2) \right )
\end{equation}
on the left segment and by

\begin{equation} 
\phi_2=\sum_n \left [ 
c_{2n}e^{i\alpha_n(x-b)}+\overline{c}_{2n}e^{-i\alpha_n(x-b)} \right ]
\sin \left (a_n
(y+a/2) \right )
\end{equation}
on the right segment. Here $a_n=n\pi/a, c_m=m\pi/c$,   

\begin{equation}
\alpha_n=\sqrt{k_1^2-a_n^2}
\ \ , \ \ \ \ \ \ k_1=\omega/v_1
\end{equation}
and 

\begin{equation}
\beta_m=\sqrt{k_2^2-c_m^2}
\ \  , \ \ \ \ \ \ k_2=\omega/v_2,
\end{equation}
with $v_1$ ($v_2$) the transverse speed of sound  of the material 
in the right (left ) segment.
Inside the stub, $\phi$ must vanish at $y=d-h/2$ and $y=d+h/2$; thus the basic 
y dependence is $\sin \left (k\pi y_{\pm}/h \right )$, where $y_{\pm}=y\mp 
h/2-d$. 
However, the internal solution 
should also vanish at each side of the stub outside the main 
segments, and smoothly connect to the external one 
across the contact 
boundaries between the stub and the segments (at $x=0$ and $x=b$). We first 
construct two auxiliary sets of solutions to the wave equation, one of which 
matches the waveguide on the left and one on the  
right, with each vanishing elsewhere on the boundary. 
The appropriate boundary conditions are

\begin{eqnarray}
&&\chi_{_{kL}}(x=b, y)=0 \\
&&\chi_{_{kL}}(x=0,y)=\left \{ 
\begin{array}{cc}
0 ,& y > c/2\\
\sin(c_k(y+c/2)), & -c/2<y<c/2\\ 
0 ,& y < -c/2   
\end{array}
\right.
\end{eqnarray}
and

\begin{eqnarray}
&&\chi_{_{kR}}(x=0, y)=0 \\
&&\chi_{_{kR}}(x=b,y)=\left \{ 
\begin{array}{cc}
0 ,& y > a/2\\
\sin(a_k(y+a/2)), & -a/2<y<a/2\\ 
0 ,& y < -a/2,   
\end{array}
\right.
\end{eqnarray}
where the subscript $k$ ($j\equiv L, R$) refers to the number of modes 
(left/right segments). The solutions $\chi_{_{kR}}$ are expanded as 

\begin{equation}
\chi_{_{kR}}=\sum_n \left [ u_n\sin (\gamma_nx)+ v_n \cos (\gamma_nx)
\right ] \sin\left (h_ny_- 
\right ),
\end{equation}
where $y_-=y+h/2-d$, $h_n=n\pi/h$, and  

\begin{equation}
\gamma_n=\sqrt{k_s^2-h_n^2}
\ \ , \ \ \ \ \ \ k_s=\omega/v_s,
\end{equation}
with $v_s$ the transverse speed of sound of the material
in the stub.
The boundary condition at $x=0$ requires $v_n=0$, whereas the condition
at $x=b$ yields

\begin{equation}
\sum_nu_n \sin(\gamma_nb)\sin(h_ny_-)
=\left \{
\begin{array}{cc}
0 ,& y > a/2\\
\sin(a_k(y+a/2)), & -a/2<y<a/2\\
0 ,& y < -a/2.
\end{array}
\right.
\end{equation}
This is a Fourier expansion with the coefficients $u_n$ given by

\begin{equation}
u_m=\frac{2}{h\sin(\gamma_mb)} I^R_{km}\ \ , \ \ \ \ m= 1, 2, ...,
\end{equation}
where

\begin{eqnarray}
I^R_{km} &=& \int^{+a/2}_{-a/2} dy \,
\sin \left (a_k(y+a/2) \right )
\sin \left (a_m
y_- \right )  \\
&=& \frac{a}{2\pi} \left \{ \frac{1}{(k-ma/h
)} \left [
\sin \left ( k\pi-h_m
s_+\right )
+\sin \left ( h_ms_-\right ) \right ] 
\right.\nonumber \\
&&\hspace*{2mm}-\left.\frac{1}{(k+ma/h
)} \left [
\sin \left ( k\pi+h_m s_+\right )
-\sin \left ( h_ms_-\right )
\right ] \right \} 
\hspace*{8mm} 
\end{eqnarray}
and $s_{\pm}=h/2\pm a/2-d$; then  from Eq. (15) we have 

\begin{equation}
\chi_{kR}=\frac{2}{h}\sum_m\ \frac{\sin(\gamma_mx)}{\sin(\gamma_mb)}
\ I^R_{km}\sin \left (h_my_- \right )
\end{equation}
Following the same procedure, we find that 

\begin{equation}
\chi_{kL}=\frac{2}{h}\sum_m\ \frac{\sin(\gamma_m(b-x))}{\sin(\gamma_mb)}
\ I^L_{km}\sin \left (h_my_- \right )
\end{equation}
where $I^L_{km}$ is defined just as $I^R_{km}$ with $a$ replaced by $c$. The 
actual wave function in the stub region can be expanded in terms of these 
auxiliary solutions $\chi_{_{kR}}$ and $\chi_{_{kL}}$ 

\begin{equation} 
\phi_s=\sum_k \left ( f_k\chi_{_{kL}}+\overline{f}_k\chi_{_{kR}}\right ).
\end{equation}
The continuity of the wave function at $x=0$ and $b$ requires  
$f_k=c_{1k}+\overline{c}_{1k}$ and $\overline{f}_k=c_{2k}+\overline{c}_{2k}$.
Thus one can write

\begin{equation}
\phi_s=\frac{2}{h}\sum_{km}\left [
(c_{2k}+\overline{c}_{2k})\ \frac{\sin (\gamma_mx)}{\sin (\gamma_mb)}\ I^R_{km}
+(c_{1k}+\overline{c}_{1k})\ \frac{\sin (\gamma_m(b-x))}{\sin (\gamma_mb)}\ 
I^L_{km}
\right ] \sin\left (h_m
y_- \right  ).
\end{equation}
Similarly, matching the derivative at $x=0$ gives

\begin{eqnarray}
\sum_n(c_{1n}-\overline{c}_{1n})\,i\beta_n \sin \left (c_n
(y+c/2) 
\right )\hspace*{+9cm}\nonumber \\
=\frac{2}{h}\sum_{km}\frac{(c_{2k}+\overline{c}_{2k})I^R_{km}-
(c_{1k}+\overline{c}_{1k})\cos (\gamma_mb)I^L_{km}}{\sin (\gamma_mb)}
\, \gamma_m \sin \left (h_my_- \right ),\hspace*{+0.8cm} 
\end{eqnarray}
and at $x=b$,

\begin{eqnarray}
\sum_n(c_{2n}-\overline{c}_{2n})\,i\alpha_n \sin \left (a_n(y+a/2) 
\right )\hspace*{+9cm}\nonumber \\
=\frac{2}{h}\sum_{km}\frac{(c_{2k}+\overline{c}_{2k})\cos(\gamma_mb)I^R_{km}-
(c_{1k}+\overline{c}_{1k})I^L_{km}}{\sin (\gamma_mb)}
\, \gamma_m \sin \left (h_my_- \right ).\hspace*{+0.8cm} 
\end{eqnarray}
Multiplying Eq. (27) by $\sin(c_l(y+c/2))$ on both sides and integrating 
from $-c/2$ to $c/2$ gives

\begin{equation}
(c_{1l}-\overline{c}_{1l})i\beta_l=\frac{4}{ch}\sum_{km}
\frac{(c_{2k}+\overline{c}_{2k})I^R_{km}-
(c_{1k}+\overline{c}_{1k})\cos(\gamma_mb)I^L_{km}}{\sin(\gamma_mb)}\,
\gamma_mI^L_{lm}.
\end{equation}
Similarly, multiplying Eq. (28) by $\sin(a_l(y+a/2))$ on both sides and 
integrating from $-a/2$ to $a/2$ yields

\begin{equation}
(c_{2l}-\overline{c}_{2l})i\alpha_l=\frac{4}{ah}\sum_{km}
\frac{(c_{2k}+\overline{c}_{2k})\cos(\gamma_mb)I^R_{km}-
(c_{1k}+\overline{c}_{1k})I^L_{km}}{\sin(\gamma_mb)}\,\gamma_mI^R_{lm}.
\end{equation} 
We define

\begin{equation}
c^{\pm}_k=c_k \pm \overline{c}_k 
\end{equation}
to cast Eqs. (29) and (30) in the form

\begin{equation}
i\beta_lc^-_{1l}=\frac{4}{ch}\sum_{km}\frac{\gamma_mI^L_{lm}}
{\sin(\gamma_mb)} \left [ c^+_{2k}I^R_{km}-
c^+_{1k}\cos(\gamma_mb)I^L_{km} \right ].
\end{equation}
and

\begin{equation}
i\alpha_lc^-_{2l}=\frac{4}{ah}\sum_{km}\frac{\gamma_mI^R_{lm}}
{\sin(\gamma_mb)} \left [ c^+_{2k}\cos(\gamma_mb)I^R_{km}-
c^+_{1k}I^L_{km} \right ].
\end{equation}
We now define matrices $\hat{A}$, $\hat{B}$, $\hat{D}$, $\hat{E}$, and 
$\hat{\alpha}$ 
whose elements are

\begin{equation}
A^{LL}_{lk}=\frac{4}{ch}\sum_m\,\frac{\cos(\gamma_mb)}{\sin(\gamma_mb)}
\, \gamma_m I^L_{lm}I^L_{km}\ \ ,
\end{equation}

\begin{equation}
B^{LR}_{lk}=\frac{4}{ch}\sum_m\,\frac{1}{\sin(\gamma_mb)}
\, \gamma_m I^L_{lm}I^R_{km}\ \ ,
\end{equation}

\begin{equation}
D^{RL}_{lk}=\frac{4}{ah}\sum_m\,\frac{1}{\sin(\gamma_mb)}
\, \gamma_m I^R_{lm}I^L_{km}\ \ ,
\end{equation}

\begin{equation}
E^{RR}_{lk}=\frac{4}{ah}\sum_m\,\frac{\cos(\gamma_mb)}{\sin(\gamma_mb)}
\, \gamma_m I^R_{lm}I^R_{km}\ \ ,
\end{equation}

\begin{equation}
\alpha_{lk}=i\alpha_l \delta_{lk}\ \ , \ \ \beta_{lk}=i\beta_l \delta_{lk}. 
\end{equation}
We also define the column vectors $C^+_i$ and $C^-_i$ whose elements are 
$c^+_{ik}$ and 
$c^-_{ik}$, respectively. In this notation, we have 

\begin{equation}
\hat{\beta}C^-_1=-\hat{A}C^+_1+\hat{B}C^+_2,
\end{equation}

\begin{equation}
\hat{\alpha}C^-_2=-\hat{D}C^+_1+\hat{E}C^+_2,
\end{equation}
where $\hat{A}$, $\hat{B}$, $\hat{D}$, and $\hat{E}$ are real-valued matrices. 
These two equations determine $C^+_1$ and $C^-_1$ in terms of $C^+_2$ and 
$C^-_2$. The result is

\begin{equation}
\left (
\begin{array}{c}
C^+_1 \\
C^-_1
\end{array}
\right )=\hat{M}^{\beta \alpha}
\left (
\begin{array}{c}
C^+_2 \\
C^-_2
\end{array}
\right ),
\end{equation}
where $\hat{M}^{\beta \alpha}$ is the resulting transfer matrix with matrix 
elements

\begin{equation}
\hspace*{-1cm}M^{\beta \alpha}_{11}=\hat{D}^{-1}\hat{E}\ \ , \hspace*{3.5cm}
M^{\beta \alpha}_{12}=-\hat{D}^{-1}\hat{\alpha}
\end{equation}

\begin{equation}
M^{\beta \alpha}_{21}=-\hat{\beta}^{-1}\hat{A}\hat{D}^{-1}\hat{E}+
\hat{\beta}^{-1}\hat{B}\ \ ,\hspace*{+1cm}
M^{\beta \alpha}_{22}=\hat{\beta}^{-1}\hat{A}\hat{D}^{-1}\hat{\alpha}.
\end{equation}
This transfer matrix relates the incoming to the outgoing wave across the stub 
for arbitrary initial conditions.

The other building block of a multiple-stub system is a stubless 
waveguide segment. The transfer matrix induced by the segment of length 
$l_{ij}$ connecting the $i$th and $j$th stub is a special case of the matrix 
$\hat{M}^{\beta \alpha}$ in Eq. (41). This is obtained by considering the 
special case $c=a=h$, $d=0$, $\alpha_n=\beta_n$, and $k_2=k_1=k_s$. The result is

\begin{equation}
\hat{P} (ij)= \left [
\begin{array}{cc}
\cos(\alpha l_{ij})  & -i\sin(\alpha l_{ij}) \\
-i\sin(\alpha l_{ij}) &  \cos(\alpha l_{ij})
\end{array}
\right ],
\end{equation}
where $j=i\pm 1$. Given $\hat{M}^{\beta \alpha}$ and 
$\hat{P}$, the total transfer matrix for an n-stub system is 

\begin{equation}
\hat{M}^T=\prod^{n-1}_{i=1} \left [ \hat{M}^{\beta \alpha}(i) \hat{P}(i,i+1)    
\right ]\hat{M}^{\beta \alpha}(n)
\end{equation}
Suppose the incident wave, at the entrance of the first stub, is represented by 
$\{C^+_{in}, C^-_{in} \}$ and the outgoing one, at the exit of the last stub, by 
$\{C^+_{out}, C^-_{out} \}$. Then

\begin{equation}
\left ( 
\begin{array}{c}
C^+_{in}\\
C^-_{in}
\end{array}
\right )
=\hat{M}^T
\left (
\begin{array}{c}
C^+_{out}\\
C^-_{out}
\end{array}
\right ).
\end{equation}
We now discuss the physical conditions imposed on the incoming and outgoing 
wave components. Depending on the frequency, $\alpha_n$ and $\beta_n$ could 
be real or pure imaginary. We choose them to be positive for an open 
channel and lying on the positive imaginary axis for a closed channel,
so a complete set of eigenfunctions of Eq. (8) is obtained. With
this choice, $\overline{C}_{out,n}$ represents either a leftward moving wave or 
an 
exponentially divergent wave at positive infinity. Now, for the sake of 
the argument, we will take the incident wave to come from the left; in this
situation, we must set $\overline{C}_{out}=0$ and thus 
$C^+_{out}=C^-_{out}=C_{out}$. Physically $c_{out,n}$ represents the transmitted 
wave components in the $n$th mode. In the entrance region the components 
$\overline{c}_{in,n}$ are allowed. Here 
$\overline{c}_{in,n}$ represents the amplitude of the reflected wave in the 
$n$th mode in the open channel. For a closed channel the reflected wave is a 
transient and decays exponentially. Explicitly,

\begin{equation}
C^+_{in}=C_{in}+\overline{C}_{in}=\hat{M}^T_{11}C_{out}+\hat{M}^T_{12}C_{out},
\end{equation}

\begin{equation}
C^-_{in}=C_{in}-\overline{C}_{in}=\hat{M}^T_{21}C_{out}+\hat{M}^T_{22}C_{out},
\end{equation}
where $\hat{M}^T_{ij}$ are blockwise submatrices of $\hat{M}^T$. Adding these 
two equations we can determine the transmitted amplitudes $C_{out}$ by 
solving

\begin{equation}
2C_{in}=\left ( \sum_{ij}\hat{M}^T_{ij} \right ) C_{out}.
\end{equation}
Following this, the reflection coefficients are given by

\begin{equation}
\overline{C}_{in}=\frac{1}{2} \left ( 
\hat{M}^T_{11} + \hat{M}^T_{12} -\hat{M}^T_{21}-\hat{M}^T_{22} 
\right )C_{out}.
\end{equation}
The total transmission and reflection coefficients are then given by

\begin{equation}
T=\frac{\sum_{_{n\in \{open\}}} c_{out,n}c^*_{out,n}\alpha_n}
{\sum_{_{n\in \{open\}}} c_{in,n}c^*_{in,n}\alpha_n}
\end{equation}
and

\begin{equation}
R=\frac{\sum_{_{n\in \{open\}}} 
\overline{c}_{in,n}\overline{c}^*_{in,n}\alpha_n}
{\sum_{_{n\in \{open\}}} c_{in,n}c^*_{in,n}\alpha_n}.
\end{equation}
For the sake of completeness, it is noteworthy that for the bound-state 
calculation, one must solve the homogeneous version of Eq. (49).

We also stress that in this formalism, once we know how to handle the single 
stub case, the multiple-stub problem can be dealt with  little additional 
labor. This is not the case with either the recursive Green-function method 
or the usual mode-matching approach.

Finally, the band structure for a periodic system of acoustic stub tuners is 
computed by solving the standard eigenvalue equation

\begin{equation}
\hat{M}^B \Phi=e^{ik_xL}\hat{I}\Phi,
\end{equation}
where $k_x$ is the Bloch vector and $L=b+l$ is the period of the 1D superlattice 
of acoustic stub tuners. The matrix $\hat{M}^B$ is the same as $\hat{M}^T$ 
for the unit cell but 
with $c=a$ and $\beta_n=\alpha_n$, $\hat{I}$ is the unit matrix of the same 
order as $\hat{M}^B$, and $\Phi$ is the column eigenvector. So the strategy of 
the computation is that we input the dimensionless frequency 
$\Omega  =\omega L/\pi v_{wg}$ and calculate $w=\exp{(ik_xL)}$, yielding 
$k_x=-(i/L)\ln (w)$. For $\left |w \right |=1$ ($\neq 1$) one obtains bands (gaps) in the band 
structure for a given set of material and geometrical parameters. It should be 
pointed out that in order to make this paper as  self-consistent as 
possible we have heavily relied on Ref. \cite{wu} in this section.

\section{ILLUSTRATIVE EXAMPLES}

For the sake of clarity we discuss the numerical results in two parts. First, 
we consider the case when the waveguide and the stubs are made up of the same 
material. Clearly the band structure and/or 
transmission spectrum in this case reveals the influence of the various
parameters involved in the problem. Then we take up the case when the 
 materials in the waveguide and the stubs are different. Practically speaking, 
this case is more complex but richer than the previous one in the sense that one has 
more options to modulate 
the band structure and/or 
transmission spectrum. We have chosen carbon and epoxy resin as the suitable 
materials the acoustic system considered is made of. This is because these are 
the materials whose combination was first demonstrated to give rise to a 
complete band gaps, i.e., independent of the polarization of the wave and of the 
direction 
of propagation, in   2D periodic phononic crystals\cite{vas}. The 
parameters used are $\rho=1.75$ (1.2) g/cm$^{3}$ and $v_t=711,095$ (115,830) 
cm/sec for carbon (epoxy).

\subsection{Same material in waveguide and stubs}

The left part of Fig. 3 shows the first nine bands for a symmetrically stubbed 
system made up of epoxy, with  parameters  $a_L=a/L=0.5114$, 
$b_L=b/L=0.4886$, $h_L=h/L=1.125$, and $d_L=d/L=0.0$. As one can see, all 
the nine bands are separated from each other by stop bands, or gaps, within which 
the acoustic wave propagation is forbidden. Unlike the other 2D and 3D periodic 
systems\cite{kush,kush1}, 
there is a complete gap below a cutoff frequency 
$\Omega_c\simeq 1.6$ down to $\Omega=0$. It is found that this (the 
lowest) gap persists independent of the values of the variable parameters. The 
existence of all  nine gaps  is  well 
corroborated by the energy dependence of the transmission coefficient for 
$n_{stub}=50$ on the right part of Fig. 3. The numerical results clearly reveal 
the zeros and ones in the transmission. It is noteworthy that the band structure 
in this figure contains both direct and indirect gaps. 
For instance, the second, third, fifth, sixth, and ninth gaps are 
direct, while the rest are indirect. We consider it more appropriate to include 
the lowest (and also the widest) gap in the category of direct gaps. The 
most important aspect of these results is the cutoff frequency $\Omega_c$ below 
which no propagation at all is allowed.  

Figure 4 depicts the dimensionless gap widths of the three lowest gaps of
 Fig. 3 as a function of the dimensionless stub width $b_L$ and the stubs are 
symmetric, i.e., $d=0$.  The width of the lowest gap 
$\Delta_1$ decreases gradually, by approximately 45\%, with increasing $b_L$
but still remains finite for $b\rightarrow L$. 
We notice that the
maximum of $\Delta_1$ is near  the cutoff frequency of the waveguide
segments $\Delta_w=1/a_L=1.96$ and the minimum of $\Delta_1$ is near the
cutoff frequency of the stub segments $\Delta_s=1/h_L=0.89$. In general, 
the cutoff
frequency of the combined system is between the one of the waveguide segments
and that for the stub segments.
The second lowest gap
$\Delta_2$ reaches a maximum for $b_L\simeq 0.7$ and decreases slightly 
before approaching the final minimum at $b\rightarrow L$. 
As can be seen, this gap increases enormously relative to its value for zero 
stub width. Similarly, the  third lowest gap 
$\Delta_3$ reaches one maximum at $b_L\simeq 0.44$, it then vanishes at
$b_L\simeq 0.68$, and  finally reaches a maximum at $b_L\rightarrow 1$. Note 
that all three gaps start opening up at a vanishingly small but finite value of 
$b_L$ and that the lowest gap remains the widest one over the whole range of the 
stub width.

Figure 5 represents the band structure and the corresponding transmission 
spectrum for asymmetric stubs, with $d=0.25$. The rest of the 
parameters are the same as in Fig. 3. We observe that 
the asymmetry has introduced two important effects. First, the 
number of bands accommodated within the same frequency range
has increased, from nine to fourteen. 
Second, the band width of most of the bands has reduced. Overall, the effect of 
the asymmetry seems to result in gaps larger in number but shorter in width.
This is true despite some exceptions, for instance, the case of the lowest gap, 
which now extends from $\Omega=0$ to $\simeq 1.7$, instead of up to 
$\Omega\simeq 1.61$ in Fig. 3. 
All  gaps in the band structure ( left part of Fig. 5) are seen to be 
well substantiated by those in the transmission spectrum (right part of Fig. 5) 
for $n=50$.

As a function of the asymmetry parameter $d$ the three lowest gaps vary very 
little, by at most 10\%, for $0\leq d\leq 0.3$. Their dependence on $b_L$ is similar to 
that shown in Fig. 4. As a function of the stub length $h_L$ their behavior, shown
 in Fig. 6, is similar to that in Fig. 4 for $\Delta_1$ and $\Delta_2$ but
 somewhat different for $\Delta_3$.  

Figure 7 shows the transmission spectrum versus reduced frequency for a 
 system made up of eleven symmetric stubs with the central (sixth) stub 
  longer ( $h_L=1.395$) and 
wider ($b_L=0.6305$) than the rest of the identical, in width and 
height, stubs. When  not identical to the other stubs, this central stub 
constitutes a defect. 
The solid (dotted) curves correspond to the presence (absence) of this
defect. One can see that there are five complete gaps in the 
spectrum within the given  frequency range. In addition, there is also a 
pseudogap, centered at $\Omega \simeq 4.35$, that corresponds to the low 
transmission or density of states. The defect introduces sharp transmission 
peaks, marked by arrows, within the first four gaps in plane analogy with the 
electronic\cite{aki} and photonic\cite{aki1} case or with that  of surface modes 
of a truncated superlattice \cite{kush2}. Another interesting consequence of 
introducing a defect in the system is 
the appearance of antiresonances\cite{aki1}, such as the one appearing in the 
fourth  band at $\Omega\simeq 3.92$. We have noted similar effects in the case of an 
asymmetric ($d\neq 0$) defect introduced in the system.

\subsection{Different materials in waveguide and stubs}

We now present numerical results for a system in which the
waveguide  is made of epoxy and the stubs of carbon. Figure 8 shows the band 
structure and transmission spectrum for symmetric stubs; the parameters 
are $a_L=0.9$, $b_L=0.9$, $h_L=1.5$, and $d_L=0.0$. We note that there 
are only seven bands accommodated in the frequency range $0\leq \Omega \leq 
20$, and every pair of bands has a full gap in between. Moreover, the lowest 
acoustic gap extends from zero to the cutoff frequency $\Omega_c\simeq 2.3$.
Some of the bands, such as the fourth and sixth, are seen to be 
almost flat and hence have vanishingly small group velocity. All  
gaps in the band structure (left panel) correspond well to those 
in the transmission spectrum for $n=50$ on the right panel. A prompt
comparison of Figs. 8 and  3 reveals that one can achieve wider gaps in the 
band structure if the segments and stubs are made up of different materials.

Figure 9 shows the three lowest gaps as a 
function of the stub width $b_L$ for the system specified in 
Fig. 8. There are several noteworthy points. First, the lowest gap is the widest 
one and the third gap is wider than the second one until $b_L\simeq 0.49$, where 
$\Delta_1=\Delta_3$. At $b_L \simeq 0.64$, the width of the lowest and second 
lowest gaps are equal, i.e., $\Delta_1=\Delta_2$. Also $\Delta_2=\Delta_3$ at 
$b_L\simeq 0.78$ and 0.94. The second lowest gap is the widest one in the range 
specified by $0.78\leq b_L \leq 0.94$. The third lowest gap vanishes  at 
$b_L\simeq 0.87$ but reappears for $b_L> 0.87$ and becomes the widest in the 
range $0.94 \leq 
b_L \leq 0.98$. Also, $\Delta_1=\Delta_3$ at $b_l\simeq 0.83$ and $0.91$. 
Moreover, $\Delta_1=\Delta_2$ at $b_L\simeq 0.96$ and $\Delta_1=\Delta_3$ at 
$b_L \simeq 0.98$.  Finally, $\Delta_1$ reaches a maximum whereas both 
$\Delta_2$ and $\Delta_3$ vanish for $b\rightarrow L$. 
Similar to Fig.
4, the minimum of $\Delta_1$ here is near the cutoff frequency
of the waveguide segments $\Delta_w=1/a_L=1.1$ and the maximum of
$\Delta_1$ is near the cutoff frequency of the stub segments
$\Delta_s=v_s/(v_1h_L)=4.1$. Again
the cutoff
frequency of the combined system is between the one of the waveguide segments
and that for the stub segments. The relative position though depends on $b_L$
and the velocity contrast and can be outside the range $(\Delta_w, \Delta_s)$.

Figure 10 depicts the band structure and transmission spectrum considered
in Fig. 8 but with asymmetric 
stubs.  The asymmetry parameter is $d_L=0.25$ and the other parameters 
the same as those used in Fig. 8. Now there are eight bands and eight gaps 
in the band structure. The lowest gap now extends from $\Omega=0.0$ to 
$\Omega\simeq 2.4$. The asymmetry is seen to have brought about a number of 
interesting effects. A larger number of bands is accommodated in the same 
frequency range but the bandwidth  is reduced. Again, the gaps in the band 
structure (left panel) correspond well to those in the transmission spectrum for 
a system with $n=50$ stubs (right panel).

As a function of the stub width $b_L$ the three lowest gaps behave 
qualitatively as those of Fig. 9. More important is their dependence, shown in Fig. 11 (a), 
on the velocity contrast between two materials. As can be seen a wide modulation can be achieved
in this asymmetric structure by just changing the ratio $v_g/v_{wg}$, more than 
two orders of magnitude for $\Delta_2$. A less pronounced variation of the same gaps
is shown in Fig. 11 (b) as a function of the stub length $h_L$. As can be seen, 
$\Delta_1$ remains almost insensitive to changes in $h_L$ and $\Delta_2$ changes by at most 
30\%; however, $\Delta_3$ can change by a factor of 10 reaching a maximum at  $h_L\sim 2.4$.  

Figure 12 shows the transmission spectrum versus  reduced frequency for a 
symmetric defect introduced in an otherwise periodic system with seven stubs 
($n=7$). The central (fourth) stub is defect in 
the sense that its length ($h_L=3.0$) and width ($b_L=0.4$) are different than 
those of the rest of the stubs. The other parameters  are $a_L=0.6$, 
$b_L=0.2$, $h_L=1.4$, $d_L=0.0$. The solid and dotted curves correspond, 
respectively, to the 
presence (absence) of the defect. There are five complete gaps in this frequency 
range in the spectrum  before introducing the defect.  Inserting this single 
defect in the system gives rise to one peak in the third gap and another in the 
fourth gap. These transmission peaks correspond to  defect modes similar to 
those appearing in Fig. 7. 

Finally, Fig. 13 illustrates the transmission spectrum versus reduced 
frequency $\Omega$ in a symmetric system made up of one (top panel), two (middle 
panel), and five (bottom panel) stubs. We remind that the waveguide segments 
(stubs) are made of epoxy (carbon) materials. The parameters are 
$a_L=b_L=h_L=0.9$. That is, the waveguide has no stubs.  The lowest panel shows the corresponding band structure.
For $n=1$ (top panel), the transmission 
coefficient  becomes very small but it never approaches zero. In this sense we
have only pseudogaps, not full gaps, in the system. As $n$ increases the 
pseudogaps gradually turn into complete gaps (with transmission equal to zero) centered at 
almost the same midgap frequency. It has also been observed that the number of such 
complete gaps increases with increasing $n$.

\section{CONCLUDING REMARKS}

We have investigated the existence of tunability of complete spectral gaps in 
the band structure of a quasi-one-dimensional waveguide with double stubs
periodically grafted at $N$ equidistant sites. The waveguide segments and the 
stubs can be made up of same or different 
materials. The latter case is found to be relatively more interesting since 
there one has more and better options to achieve the complete gaps. The single 
symmetric defect is shown to introduce extra modes
in the gaps of an otherwise periodic system that result in transmission 
resonance peaks. While the computation of the band structure 
requires an infinitely long periodic system, the 
transmission spectrum is calculated only for a finite system. As shown though, 
the gaps in the band structure correspond well to those
in the transmission spectrum. Accordingly, we conclude that the 
transmission spectrum in all cases remains consistent for 
$n\geq 10$.

The numerical results we presented  pertain  only to the out-of-plane modes. 
However, as mentioned already in Sec. II, A, the equations describing the in-plane modes 
have the same structure in a homogeneous medium. Explicitly, the transverse in-plane modes have exactly the same 
band structure as the out-of-plane modes. As for the longitudinal in-plane modes,
their band structure and transmission can be 
obtained directly from Figs. 3 -- 7 by changing only the  frequency scale
when the waveguide and the stubs are made of the same material. 
If this material is different, the separation of the in-plane modes
in longitudinal and transverse is no longer possible for the structures of
Sec. III B (see Sec. II A).

The version of the transfer-matrix method employed\cite{wu} to accomplish the 
present investigation is simple and efficient\cite{aki,aki1,wu} for solving scalar 
equations. Its efficiency has already been demonstrated 
in serially connected electronic\cite{aki,wu}, optical\cite{aki1}, or the 
present acoustic devices. As stressed already in its presentation, the method  
offers important advantages in comparison to other mode-matching or recursive
Green's function techniques applied to similar problems.

In principle, the systems we studied here have potential applications in 
the designing of transducers and ultrasonic filters. We hope that the present 
findings will be tested in future experiments. 

\acknowledgements
This work was supported by the NSERC grant \# OGP0121756. The work of 
M.S.K. was also supported by CONACyT Grant \# 28110-E. 


\newpage
\begin{figure}

\caption{Schematics of a quasi-one-dimensional periodic waveguide. The double
        stubs can be made of the same or different material than that of the main
        waveguide. $L=b+l$ is the  period of the system.\\}
         
\caption{Schematics of a general unit cell with asymmetric stubs.\\} 
\ \\
 
\caption{Band structure (left panel) and transmission spectrum (right panel)
        for a system with the same material, epoxy, in the waveguide and the stubs.
        The reduced wave vector and frequency are defined by
        $k_xL/\pi$ and $\Omega=L\Omega/\pi v_1$, where 
         $v_1$ is the transverse speed of sound in the waveguide. 
        Notice  the lowest acoustic gap below the cutoff
        frequency $\Omega_c\simeq 1.6$. For the plot of the transmission
        a system of fifty ($n=50$) stubs was considered.\\}

\caption{The widths of the three lowest gaps as a function of the stub width 
	$b_L=b/L$. The solid, dashed, and dotted curves refer to the lowest ($\Delta_1$),
        second lowest ($\Delta_2$), and third lowest ($\Delta_3$) gaps. The  
	material and  the rest of the parameters are the same as those in Fig. 3.\\}

\caption{Same as in Fig. 3, but for  asymmetric stubs  with  asymmetry parameter
	 $d_L=0.25$. Notice  the lowest acoustic gap
       below the cutoff frequency $\Omega_c\simeq 1.7$.\\}
       
\caption{The widths of the three lowest gaps as a function of the stub length 
	$h_L=h/L$. The curves are marked as in Fig. 4.\\}  
\caption{Transmission spectrum for a system with eleven symmetric double
       stubs ($n=11$). The central (sixth) symmetric stub is a 
       defect with $h_L=1.395$ and $b_L=0.6305$. The rest
       of the parameters are the same as in the previous figures. Notice that the
       defect creates new modes or states in the gaps in an
       otherwise defect-free system. The peaks of these modes are marked with
       arrows.\\}
      
\caption{ Band structure (left panel) and transmission spectrum (right panel) 
	for  system with symmetric stubs with the waveguide segments (stubs) made of epoxy
        (carbon). The parameters used are $a_L=0.9$, $b_L=0.9$, and $h_L=1.5$.
       Notice the lowest acoustic gap below the cutoff frequency 
       $\Omega_c\simeq 2.3$ and extending down to $\Omega=0.0$.\\}

\caption{The widths of the three lowest gaps versus the stub width $b_L$ for the
       system studied in Fig. 8. The solid, dashed, and dotted lines refer,
       respectively, to the lowest ($\Delta_1$), second lowest ($\Delta_2$), 
       and third lowest ($\Delta_3$) gaps. Notice especially the strong variation
       of $\Delta_2$ and $\Delta_3$.\\}

\caption{Same as in Fig. 8, but for a system with asymmetric stubs ( 
	$d_L=0.25$).  The lowest acoustic gap occurs below the cutoff
       frequency $\Omega_c\simeq 2.4$ and down to $\Omega=0.0$.\\}

\caption{(a) The widths of the three lowest gaps as a function of the 
velocities' ratio $v_s/v_{wg}$. 
(b) The three lowest gaps as a function of the stub length $h_L$. 
Notice  the strong variation of $\Delta_2$ and $\Delta_3$, especially in (a). \\}

\caption{Transmission spectrum for a system with seven symmetric ($d=0$) 
	 double stubs ($n=7$). The waveguide segments (stubs) are made
        of epoxy (carbon). The central (fourth )
       stub is a  defect with $h_L=3.0$ and $b_L=0.4$. The other
       parameters are $a_L=0.6$, $b_L=0.2$, and $h_L=1.4$. The solid
       (dotted) lines refer to the transmission with (without) the defect. The
       defect gives rise to modes in the gaps in an
       otherwise defect-free system. The peaks of these modes are marked with
       arrows.\\}

\caption{ Evolution of the transmission spectrum as a function of the number of
       stubs $n$ for a system with waveguide segments (stubs) made of epoxy 
       (carbon). The parameters  are 
       $a_L=b_L=h_L=0.9$. Notice that as $n$ increases the pseudogaps 
       gradually turn into sharply defined complete gaps. The lowest panel shows 
	the  band structure. Notice that the first band is very narrow.}

\end{figure}

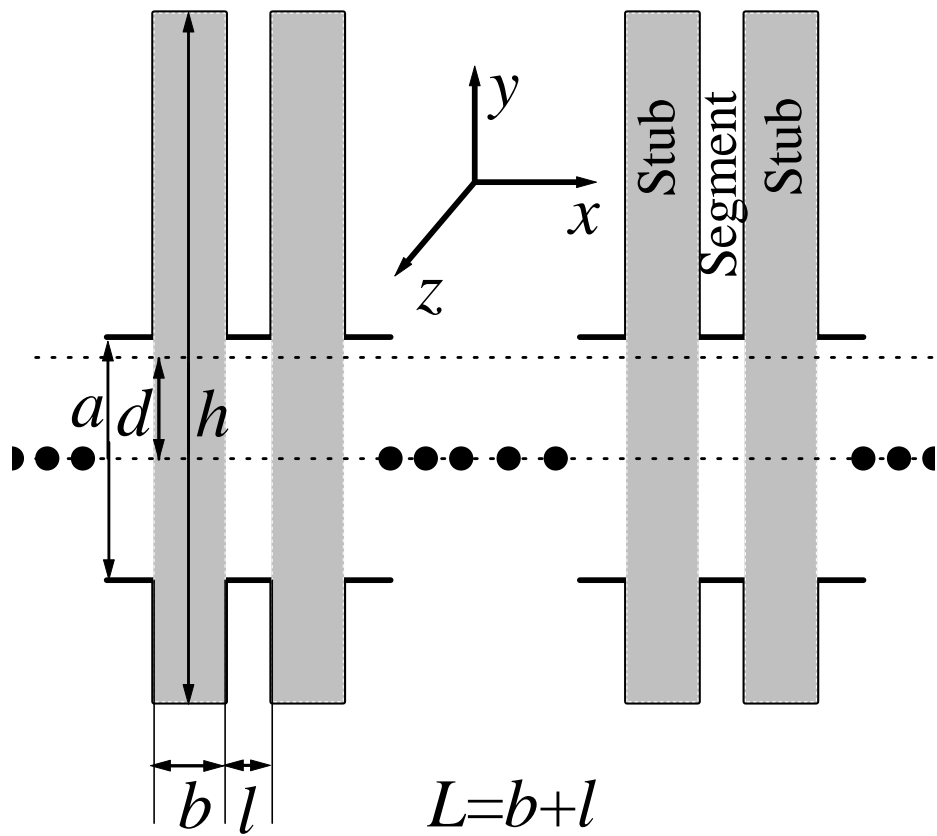
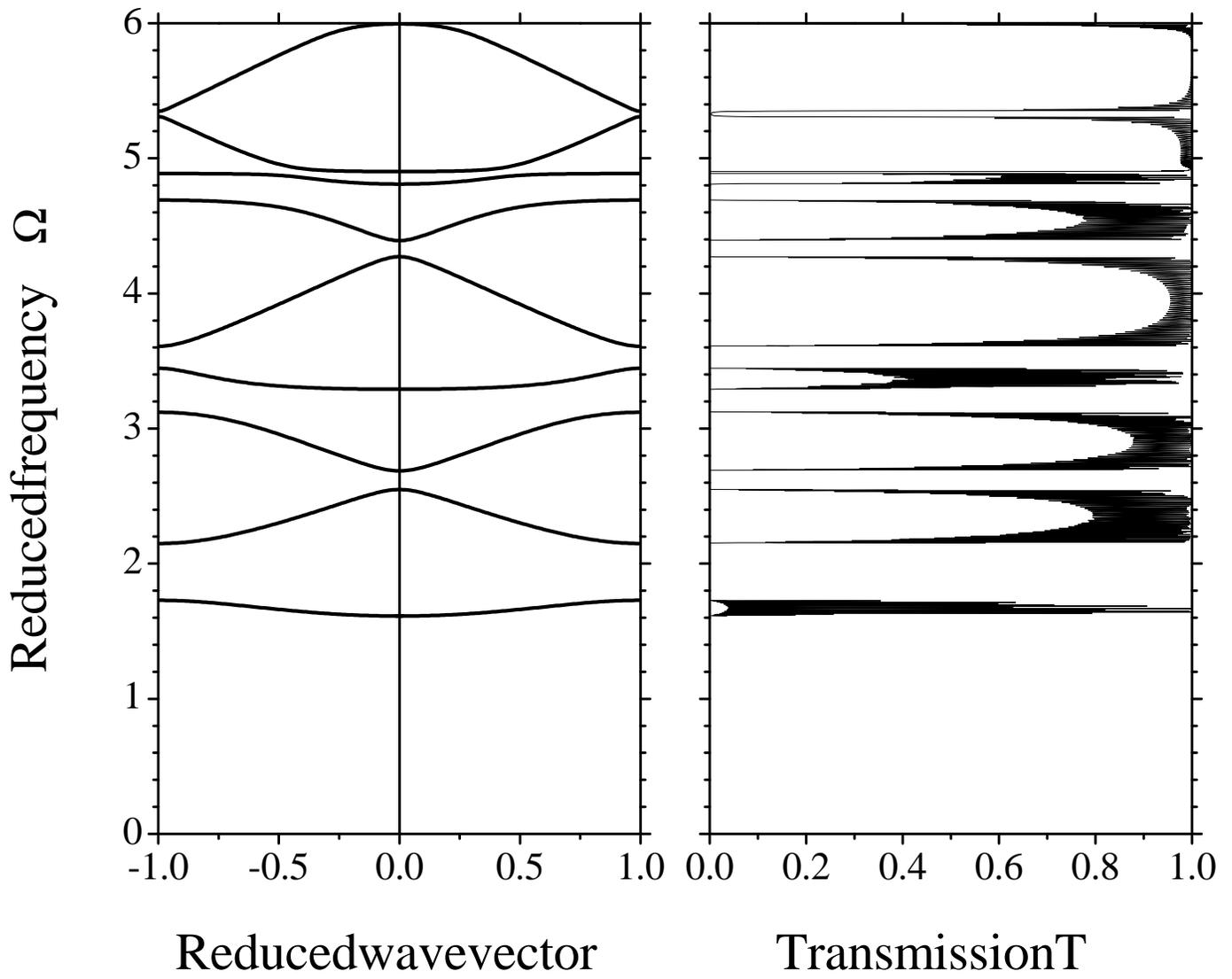
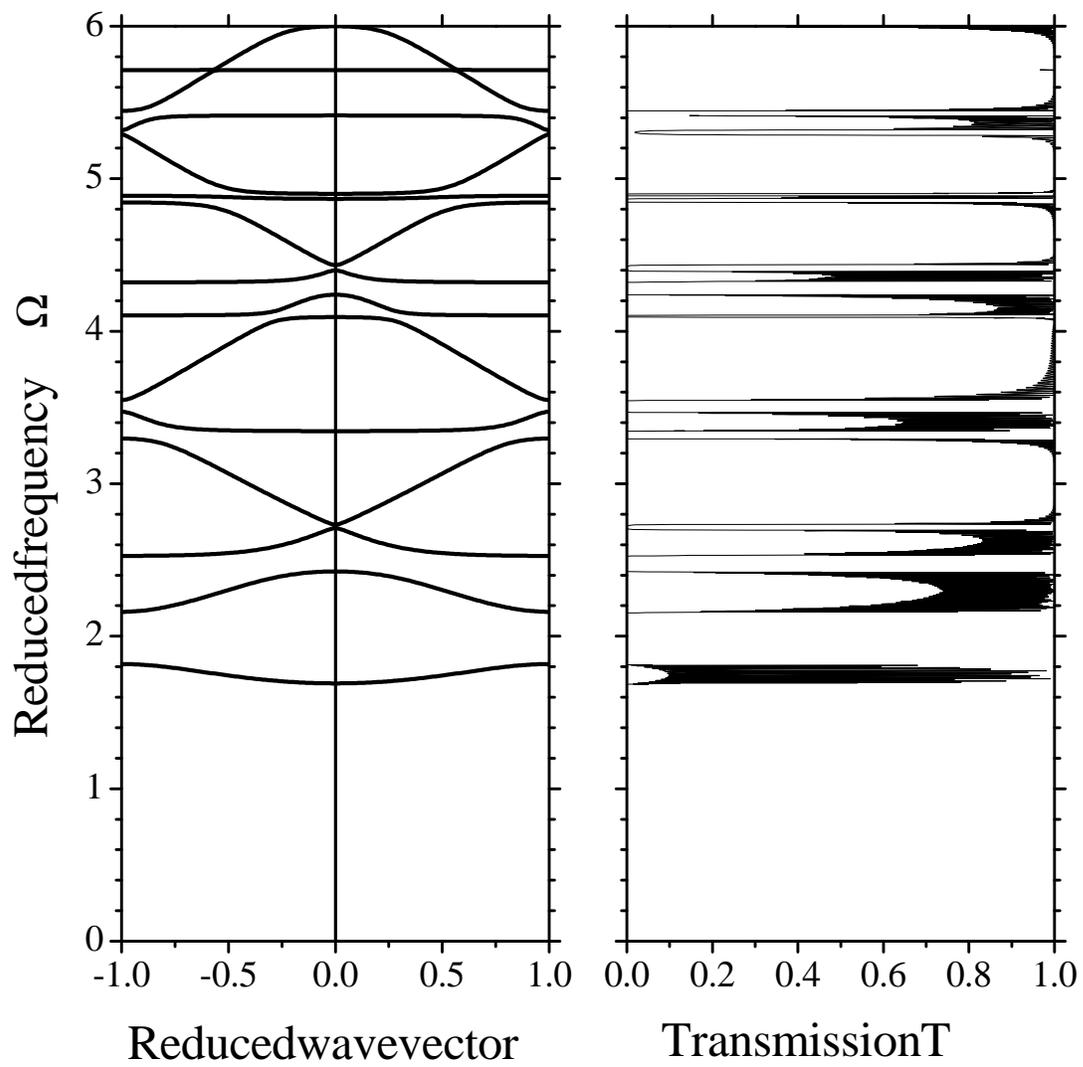
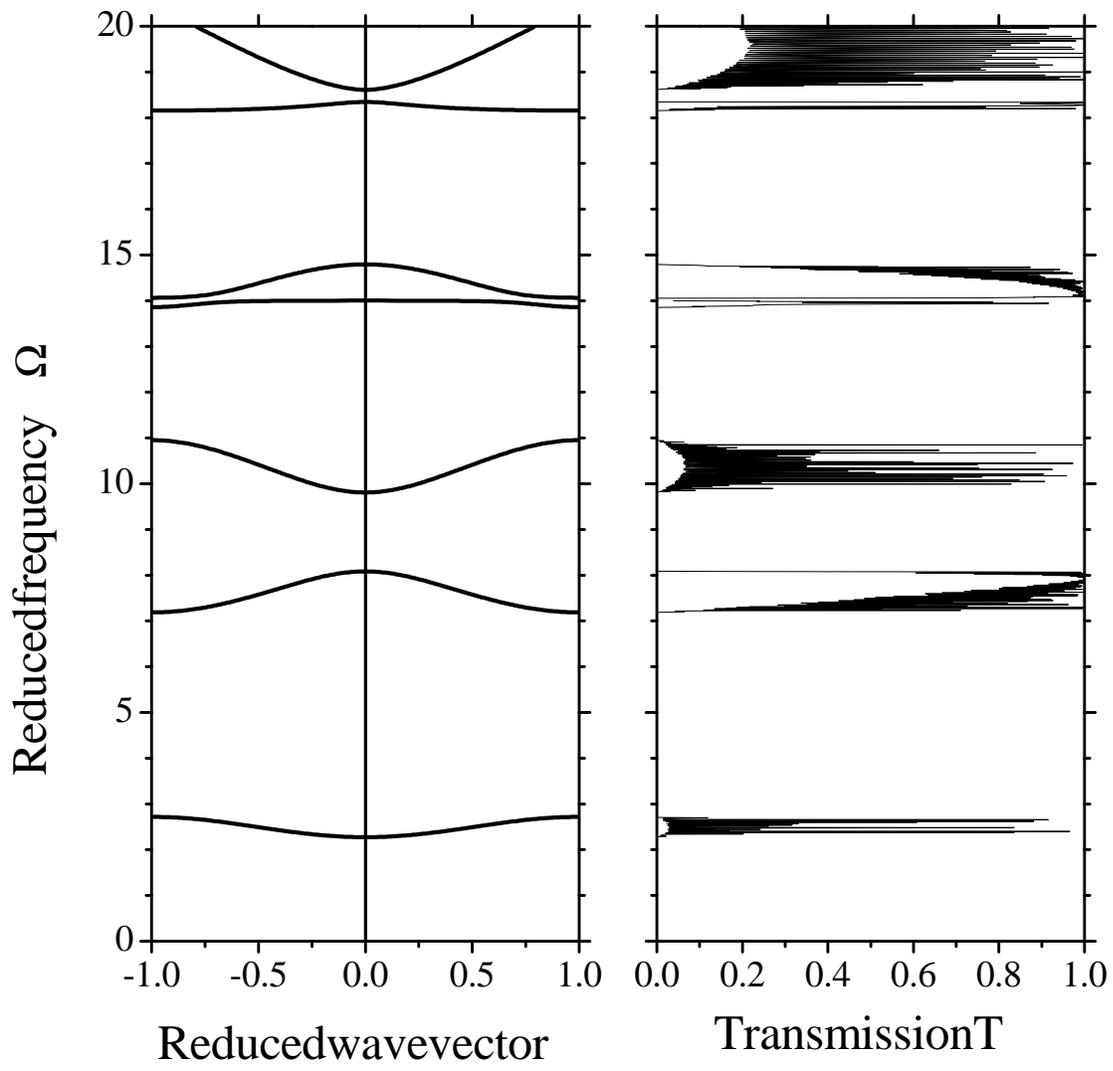
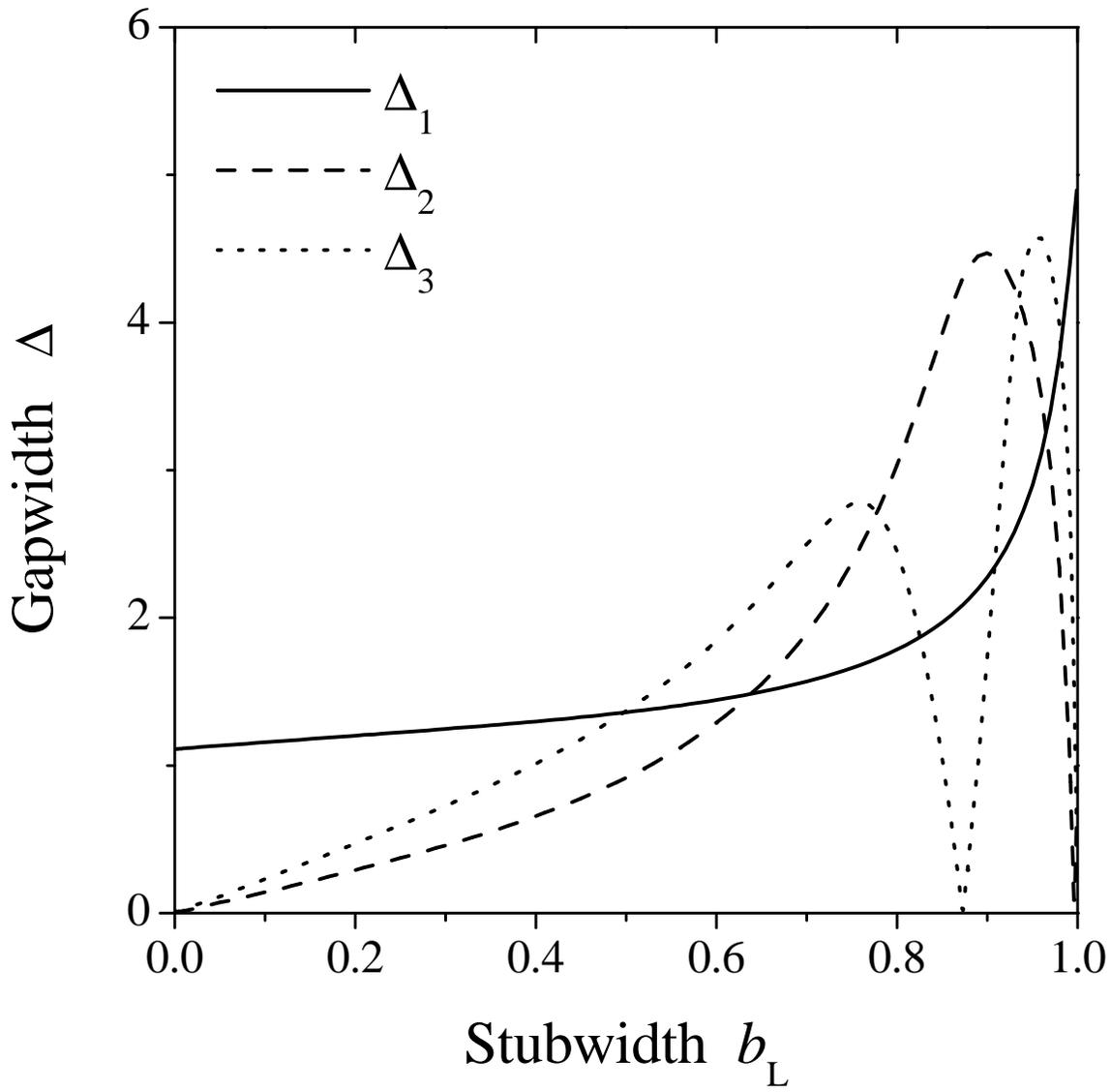
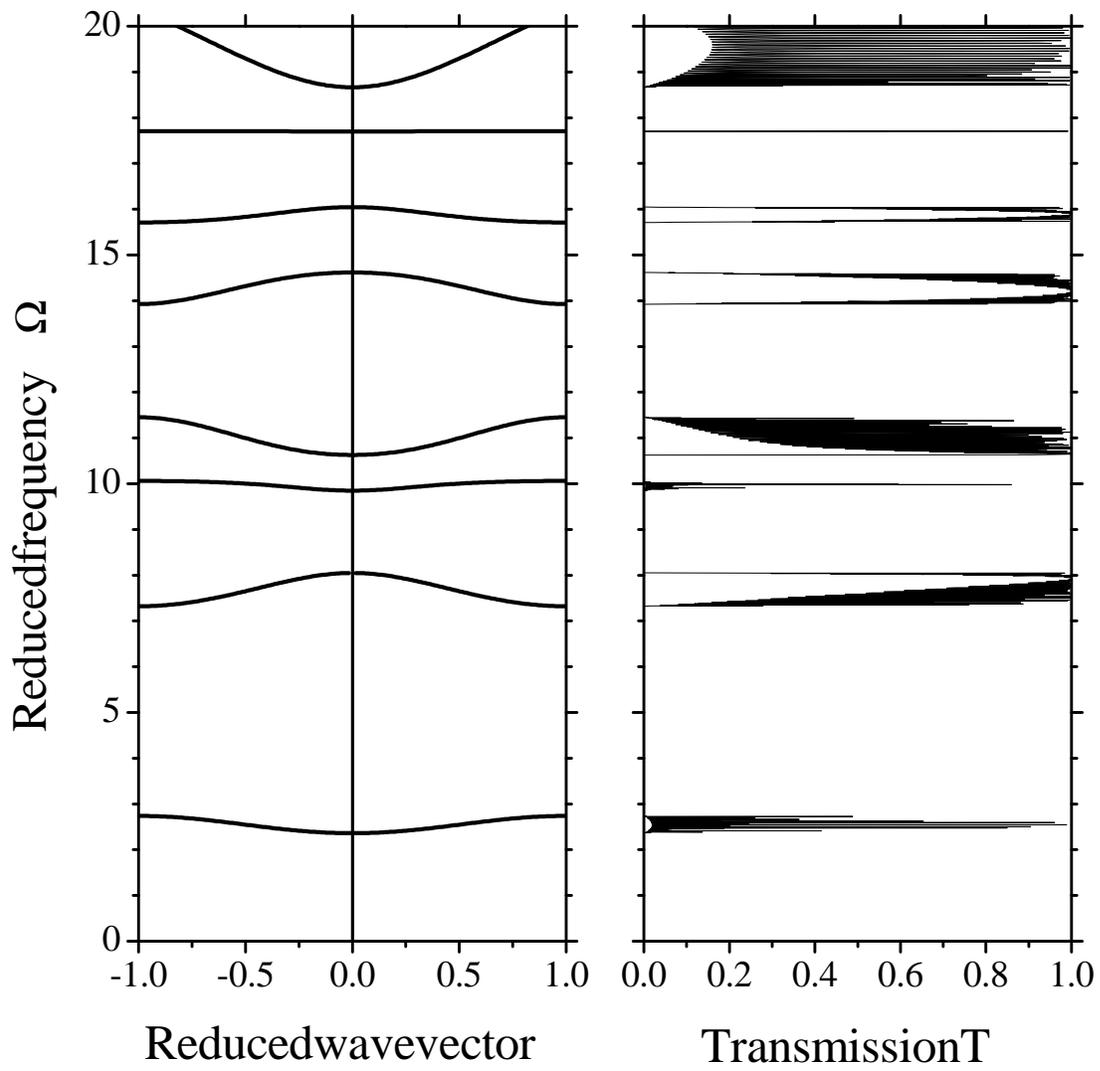
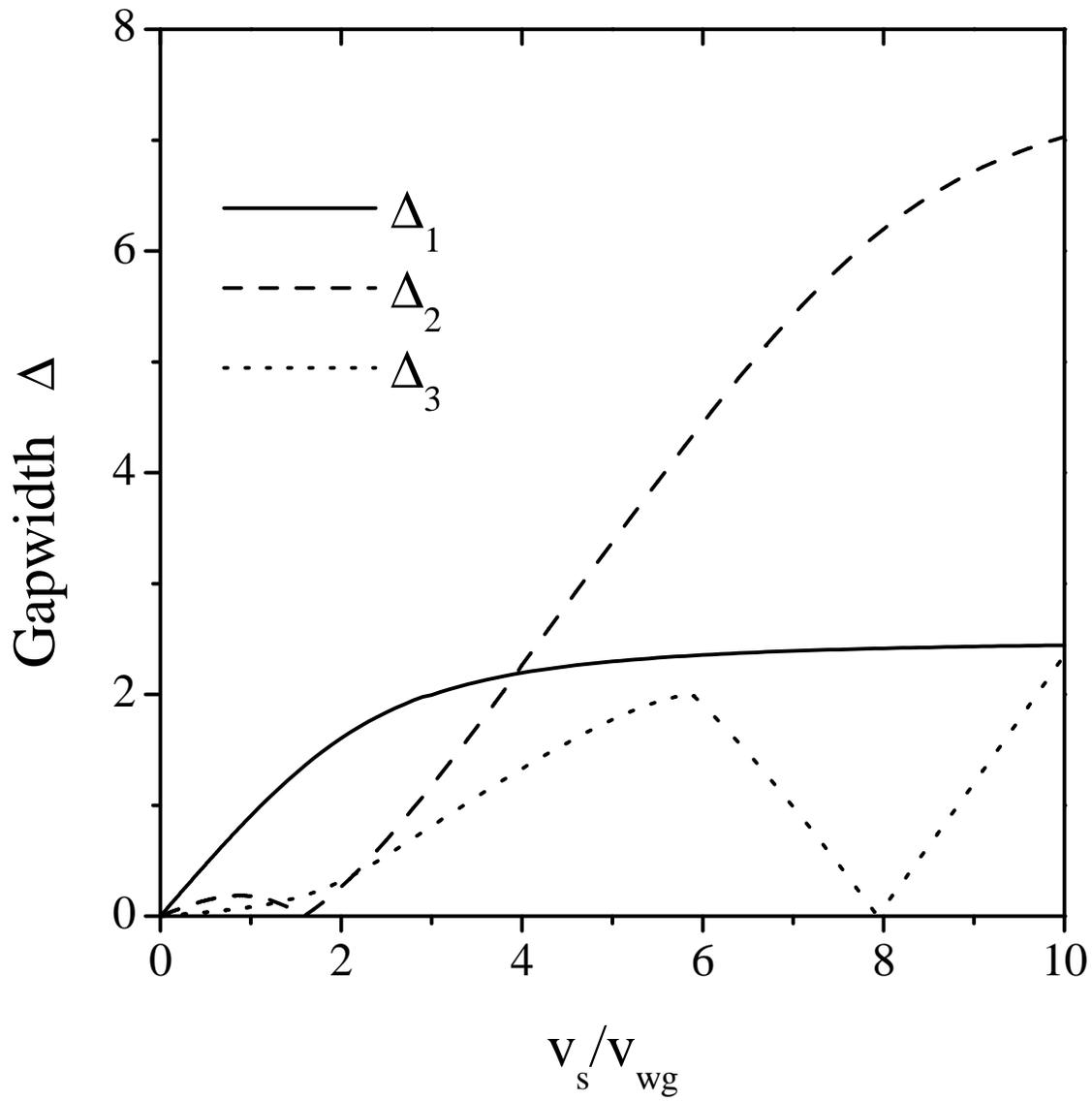
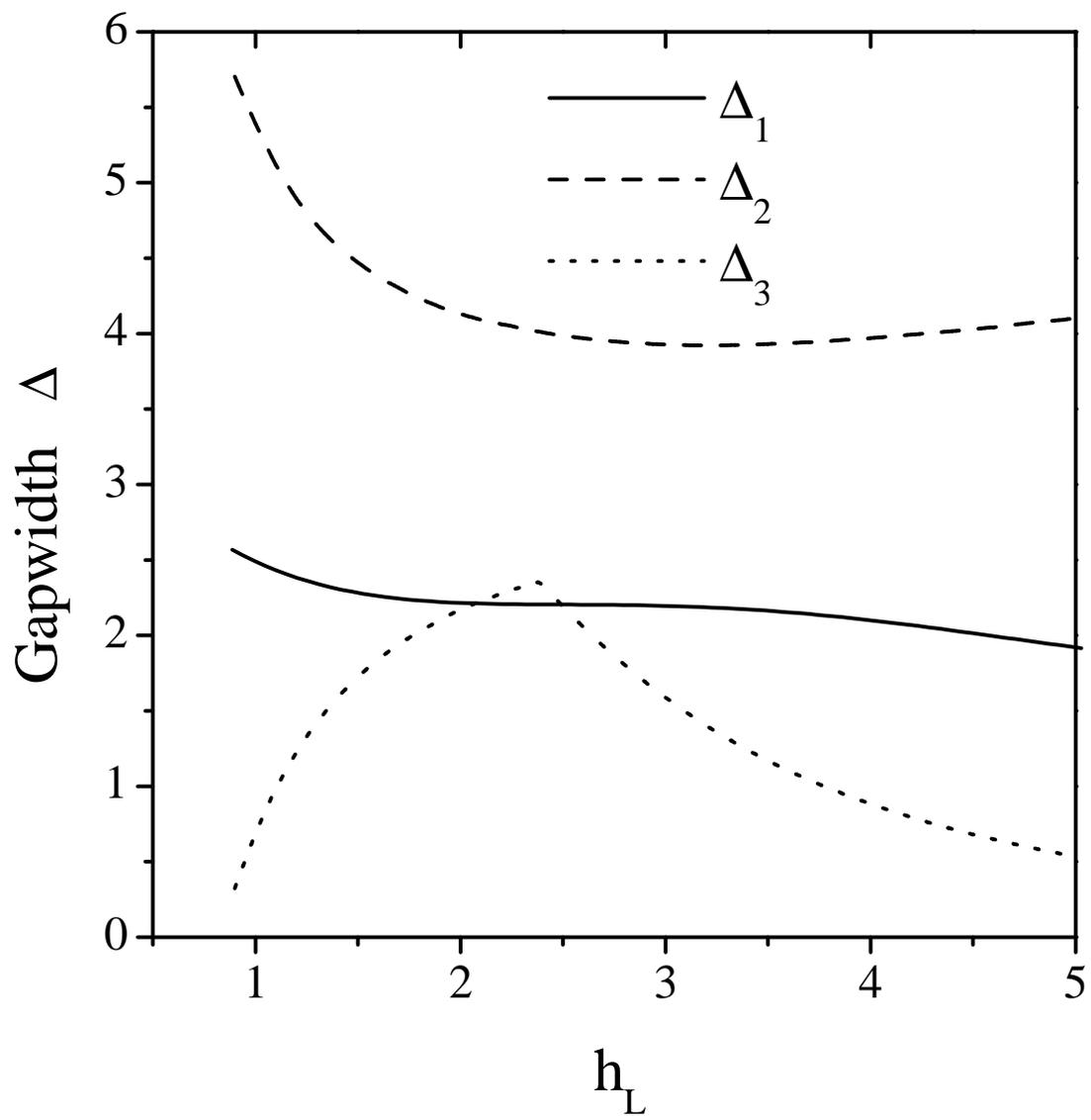
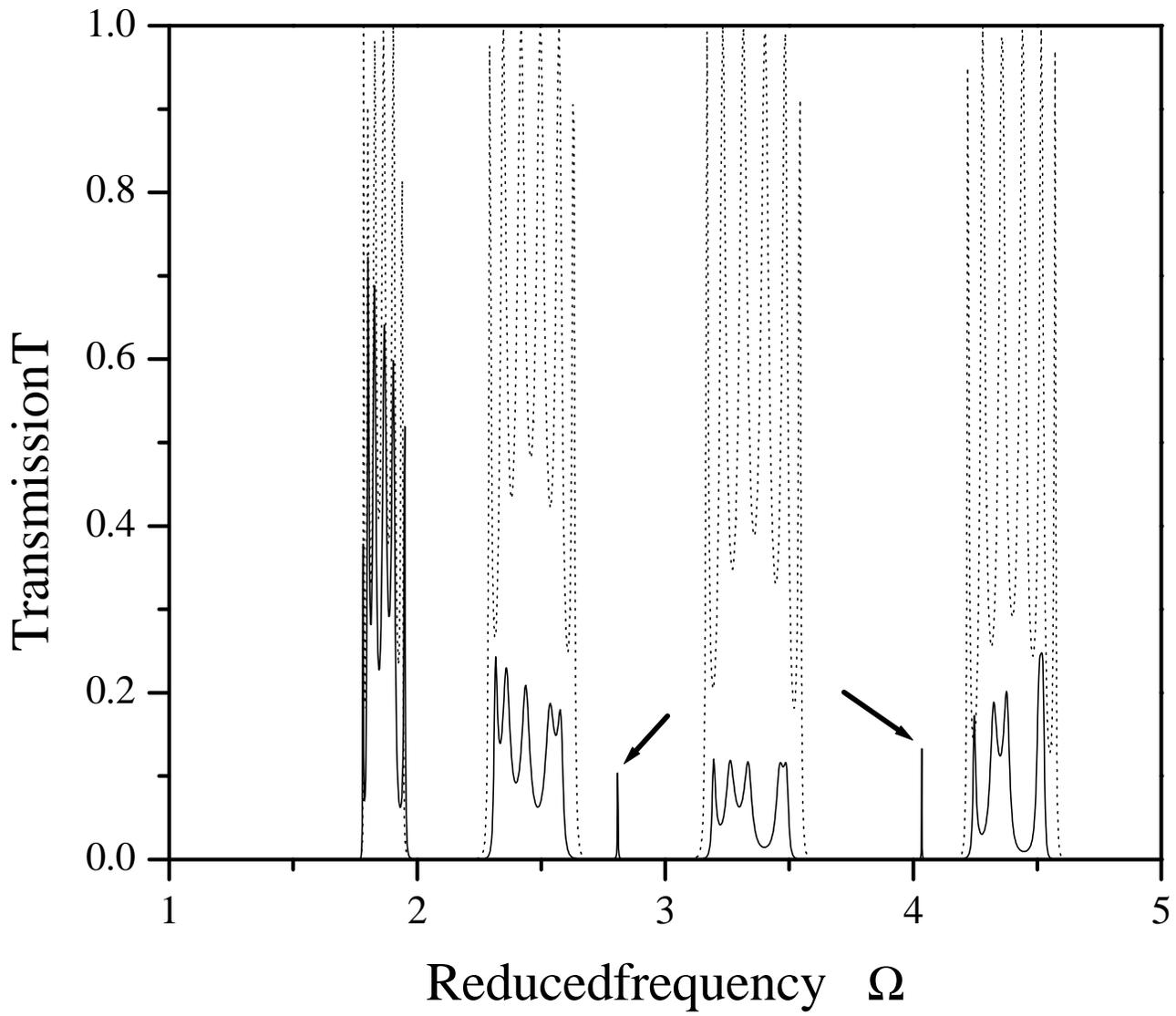
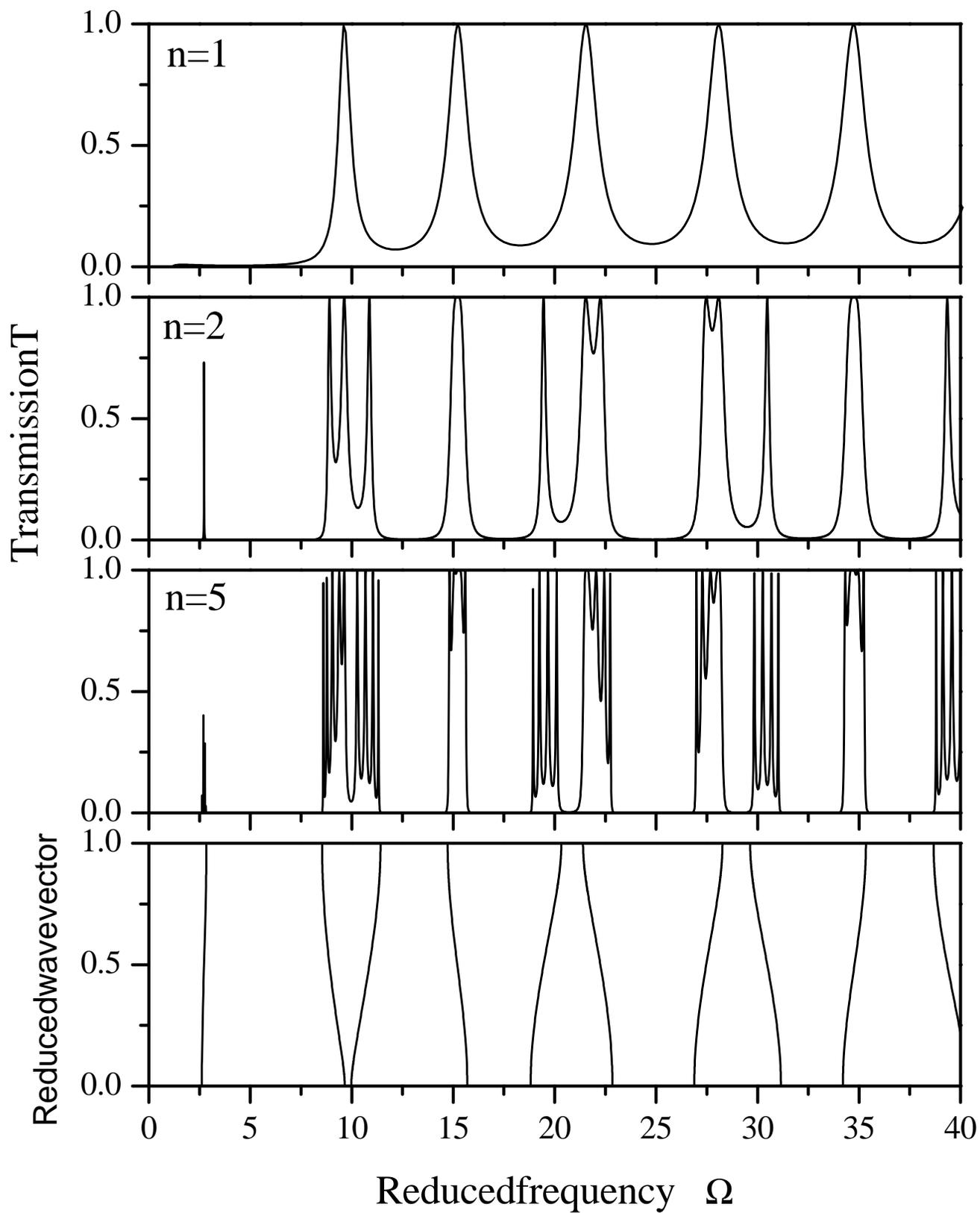
\end{document}